\documentclass[12pt]{article}
\usepackage{epsfig}
\usepackage{multicol}
\usepackage{graphicx, slashed, amsmath, amsfonts, amssymb}
\usepackage[top=1in, bottom=1in, left=1in, right=1in]{geometry}
\usepackage{rotating}
\def\bc{\begin{center}}
\def\ec{\end{center}}
\def\be{\begin{equation}}
\def\ee{\end{equation}}
\def\ba{\begin{array}}
\def\ea{\end{array}}
\def\beqn{\begin{eqnarray}}
\def\eeqn{\end{eqnarray}}
\makeatletter
\newcommand*{\rom}[1]{\expandafter\@slowromancap\romannumeral #1@}
\makeatother
 
\begin{document}
\title{Going beyond the minimal texture of quark mass matrices in the era of precision measurements}
\author{Aakriti Bagai, Shivali Kaundal, Gulsheen Ahuja*, Manmohan Gupta$^{+}$\\
{\it Department of Physics, Centre of Advanced Study,}\\
{\it Panjab University, Chandigarh, India.}\\
*gulsheen@pu.ac.in, $^{+}$mmgupta@pu.ac.in
{}}\maketitle

\begin{abstract}
An attempt has been made to carry out a detailed analysis of quark mass matrices having structure beyond the minimal texture. In particular, keeping in mind, precision measurements of CKM parameters as well as refinements in the ranges of light quark masses, we have examined the viability of all possible combinations of hermitian texture 5 zero quark mass matrices. Interestingly, one finds that all these possibilities are now excluded by the present quark mixing data, thereby, having important implications for model building.
\end{abstract}

Over the last decade, considerable progress has been made in the measurement of the quark masses, particularly in the case of the light quark masses $m_u$, $m_d$ and $m_{s}$. This has been possible primarily due to simulations based on the lattice gauge calculations \cite{flag}. Similarly, noticable progress has been made in the measuements of the Cabibbo Kobayashi Maskawa (CKM) \cite{cab, km} parameters which  can now be considered to be known at the level of `precision measurements' in the context of CKM phenomenology. In view of the relationship of the quark mass matrices with the CKM matrix, these developments regarding measurements of the quark masses as well as the CKM parameters would undoubtedly have deep implications for the structure of the mass matrices.

It is well known that in the Standard Model (SM), without loss of generality, one can always consider the quark mass matrices in the up and down sectors, $M_{U}$ and $M_{D}$, to be hermitian, with 9 free parameters in each sector. Since the number of free parameters in both the sectors is larger than the number of physical observables, i.e., 6 quark masses, 3 mixing angles and 1 CP violating phase, in order to develop viable phenomenological fermion mass matrices one has to limit this number. In this context, it may be mentioned that the idea of texture zero mass matrices \cite{fxrevart}-\cite{ourrevart2} has proved to be quite useful in explaining the fermion mixing data. One can consider a particular square mass matrix to be texture `n' zero if the sum of the number of diagonal zeros and half the number of the symmetrically placed off diagonal zeros is n. In the quark sector, texture zero mass matrices were introduced implicitly by Weinberg \cite{wein} and explicitly by Fritzsch \cite{friori1,friori2}, the original Fritzsch \emph{ans$\ddot{a}$tz} being given by 
 \begin{equation}
    M_{U}=\begin{pmatrix}
      0 & A_{U} & 0 \\
      A^{*}_{U} & 0 & B_{U} \\
      0 & B_{U}^{*} & C_{U} \\
    \end{pmatrix} ,\quad M_{D}=\begin{pmatrix}
      0 & A_{D} & 0 \\
      A^{*}_{D} & 0 & B_{D} \\
      0 & B_{D}^{*} & C_{D} \\
    \end{pmatrix},
\label{mumd}
\end{equation}
where $M_{U}$ and $M_{D}$ correspond to the mass matrices in the up (U)  and down (D) sector with complex off diagonal elements, i.e., $A_{i}=|A_{i}|e^{\text{i$\alpha $}}$ and $B_{i}=|B_{i}|e^{\text{i$\beta $}}$, where $i =U,D $, whereas $C_{i}$ is the real element of the matrix. Using the above definition of texture zero mass matrices, each of the above matrix is said to be texture 3 zero type, together these are referred as texture 6 zero quark mass matrices. 

It may be noted that since the maximum number of texture zeros which can be introduced in hermitian quark mass matrices is 3 in each sector, resulting in minimal number of parameters of the mass matrices, we refer texture 6 zero as the minimal texture of quark mass matrices. Several detailed analyses of this minimal texture of quark mass matrices have been carried out \cite{ourrevart1, ourrevart2, tsmmrrr, tsmmneelu}, of particular interest is a very recent work \cite{tex6zeroptep}, wherein an exhaustive and detailed analysis of all possible combinations of texture 6 zero quark mass matrices has been presented. Interestingly, one finds that refinements in the measurements of light quark masses as well as in the CKM matrix elements lead to all these possible combinations being ruled out even if, in future, there are reasonable changes in the ranges of the light quark masses, thereby, having important implications for model building within the top down approach. Therefore, in the absence of any viable theory of flavor dynamics, in order to arrive at a set of mass matrices which are compatible with the latest data, one needs to go for a case by case analysis of texture zero mass matrices beyond the above mentioned minimal texture. To this end, an immediate extension of the minimal texture is the texture 5 zero mass matrices. It may be noted that earlier analyses of texture 5 zero quark mass matrices \cite{tsmmrrr, tsmmneelu} show some viability with the quark mixing data. However, in these references, a comprehensive analysis indicating how and to what extent these matrices are ruled out has not been carried out. In particular, neither of the two references considers all possible texture 5 zero quark mass matrices, nor do they relate the various possibilities through permutation symmetry. This has become significant in the context
of quark lepton symmetry due to which the emphasis has now shifted to formulating the texture structure of fermion mass matrices incorporating permutation symmetry and Abelian symmetries.

Keeping these issues in mind as well as in view of the improvements in the measurements of light quark masses $m_u$, $m_d$ and $m_{s}$ as well as `precision measurements' of the CKM parameters, the purpose of the present work is to examine, in a detailed and comprehensive manner, the viability of all possible combinations of texture 5 zero hermitian mass matrices. Carrying out the present analysis of texture 5 zero mass matrices on similar lines as for texture 6 zero mass matrices \cite{tex6zeroptep}, as a first step, we have enumerated all possible texture 5 zero mass matrices and related these through permutation symmetry. Further, a detailed analysis pertaining to the viability of all these possible combinations of the mass matrices has been carried out with an emphasis to examine if, in future, there are changes in the ranges of the light quark masses, whether or not, the conclusions remain same.

In order to arrive at texture 5 zero hermitian mass matrices, one can reduce each of the texture 3 zero mass matrices mentioned in equation (\ref{mumd}), to texture 2 zero matrices by replacing either one of the diagonal zeros or two symmetrically placed off diagonal zeros by a non-zero real or complex element of the mass matrix respectively. For example, considering the 2,2 element of the above matrices $M_{U}$ and $M_{D}$ to be non zero, one obtains
 \begin{equation}
    M_{U}=\begin{pmatrix}
      0 & A_{U} & 0 \\
      A^{*}_{U} & D_{U} & B_{U} \\
      0 & B_{U}^{*} & C_{U} \\
    \end{pmatrix} ,\quad M_{D}=\begin{pmatrix}
      0 & A_{D} & 0 \\
      A^{*}_{D} & D_{D} & B_{D} \\
      0 & B_{D}^{*} & C_{D} \\
    \end{pmatrix}.
\label{mumd1}
\end{equation}
Along with these matrices, there are several other possible structures which can be considered to be texture 2 zero ones. Thereafter, one arrives at texture 5 zero mass matrices by considering either of the mass matrix in the up or the down sector to be texture 3 zero type, given in equation (\ref{mumd}), along with the mass matrix in the other sector being 2 zero type, given in equation (\ref{mumd1}). 

Coming to the total number of texture 2 zero structures, one can make use of the fact that the total number of structures for a texture `n' zero mass matrix is $^6C_n = \frac{6!}{n!(6-n)!}, $ 6 being the number of ways to enter zeros in the mass matrices. For n=2, one gets the following 15 possible structures, ${S_1}$ to $S_{15}$, for texture 2 zero mass matrices:

 \begin{list}{(i)}
	\item Two zeros along diagonal positions:\\	
\begin{center}
$S_{1}= \begin{pmatrix}
	0	& \times & \times \\ 
	\times	&0  &  \times\\ 
	\times	&\times  & \times
	\end{pmatrix},\: S_{2}= \begin{pmatrix}
	0	& \times & \times \\ 
	\times	&\times  &  \times\\ 
	\times	&\times  & 0
\end{pmatrix},\:
S_{3}= \begin{pmatrix}
\times	& \times & \times \\ 
\times	& 0  & \times\\ 
\times	& \times  & 0
\end{pmatrix},$\\ 
\end{center}	
where $\times$'s represent the non-vanishing entries.
\end{list}

\begin{list}{(ii)}
 \item One zero along diagonal position and two zeros symmetrically placed at off diagonal positions:\\
 \begin{center}
	$S_{4}= \begin{pmatrix}
	0	& 0 & \times \\ 
	0	&\times  &  \times\\ 
	\times	&\times  & \times
	\end{pmatrix},\: S_{5}= \begin{pmatrix}
	0	& \times & 0 \\ 
	\times	& \times  &  \times\\ 
	0	& \times  & \times
\end{pmatrix},\:
S_{6}= \begin{pmatrix}
0	& \times & \times \\ 
\times	& \times  & 0\\ 
\times	& 0  & \times
\end{pmatrix},$\\
$S_{7}= \begin{pmatrix}
\times	& 0 & \times \\ 
0	& 0  & \times\\ 
\times	& \times  & \times
\end{pmatrix},\:S_{8}= \begin{pmatrix}
\times	& \times & 0 \\ 
\times	&0  & \times\\ 
0	& \times  & \times
\end{pmatrix},\: S_{9}= \begin{pmatrix}
\times	& \times & \times \\ 
\times	& 0  & 0\\ 
\times	& 0  & \times
\end{pmatrix},$

$S_{10}= \begin{pmatrix}
\times	& 0 & \times \\ 
0	& \times  & \times\\ 
\times	& \times & 0
\end{pmatrix},\: S_{11}= \begin{pmatrix}
\times	& \times & 0 \\ 
\times	&\times  &\times\\ 
0	& \times  & 0
\end{pmatrix}, \:S_{12}= \begin{pmatrix}
\times	& \times & \times \\ 
\times	& \times  & 0\\ 
\times	& 0  & 0
\end{pmatrix}.$
\end{center}
\end{list}
\begin{list}{(iii)}
	\item Two zeros symmetrically placed at two off diagonal positions:\\
	\begin{center}
		$ S_{13}= \begin{pmatrix}
			\times	& 0 & 0 \\ 
			0	& \times  & \times\\ 
			0	& \times  & \times
		\end{pmatrix},\: S_{14}= \begin{pmatrix}
			\times	& 0 & \times \\ 
			0	& \times &  0\\ 
			\times	& 0  & \times
		\end{pmatrix},\:
		S_{15}= \begin{pmatrix}
		\times	& \times & 0 \\ 
			\times	&\times  &  0\\ 
			0	& 0  & \times
		\end{pmatrix}.$\\
		
	\end{center}
	\end{list}
   
As mentioned earlier, one can obtain texture 5 zero mass matrices by considering $M_{U}$ to be texture 3 zero type and $M_{D}$ to be  a texture 2 zero structure and vice versa. It may be noted that in Ref. \cite{tex6zeroptep} it has been explicitly shown that the total number of possible texture 3 zero structures to be considered as $M_{U}$ and $M_{D}$ are 13, out of which 12 are related through permutation symmetry, 6 placed in class I of Table 1 and the other 6 in class II of the table, referred to as I$_a$, I$_b$, etc.. The above mentioned 15 matrices ${S_1}$ to $S_{15}$, corresponding to texture 2 zero patterns, can be deduced from the matrices presented in classes I and II, thereby relating these through permutation symmetry. To begin with, we consider the first matrix of class I, i.e, I$_a$ which can be reduced to a texture 2 zero one in 3 ways. Firstly, we consider the 2,2 element of matrix I$_a$ to be non zero, labeling it as D, leading to the matrix III$_a$ of class III of the table. The other 5 matrices of this class have been obtained through permutation symmetry. Further, we again consider  matrix I$_a$ and now assume its 1,1 element to be non zero, labeling it as E, leading to the matrix IV$_a$ of class IV of the table. Again, the other 5 matrices of this class are related through permutation symmetry. Furthermore, we now consider two symmetrically placed non zero complex elements F$e^{\text{i$\gamma $}}$ and  F$e^{\text{-i$\gamma $}}$at 1,3 and 3,1 positions of the matrix I$_a$ respectively and therefore arrive at the matrix V$_a$ and henceforth the other matrices of class V of the table. 

Similarly, one can begin with the first matrix of class II and consider its 1,1 element to be non zero leading to the matrix VI$_a$ of the the table. Interestingly, one notes that in case we assume the symmetrically placed elements at either 1,3 and 3,1 positions or 2,3 and 3,2 positions of the matrix II$_a$ to be non zero, we again get texture 2 zero structures, however, since these possibilities are already listed in class IV, therefore, these have not been mentioned again. A closer look at the table reveals that in classes IV, V and VI, the position of zeros in 3 matrices of each class is the same as in the other 3 of the same class, e.g., for the class IV, matrices IV$_a$, IV$_b$ and IV$_c$ have similar structures as those of matrices IV$_d$, IV$_f$ and IV$_e$ respectively.  Interestingly, results of our analysis, details presented later, show that the matrices having similar structures yield same results, implying exactly the same corresponding CKM matrix. Therefore, these corresponding matrices should not be counted twice implying that in all, one obtains 15 independent structures for $M_{U}$ and $M_{D}$ being texture 2 zero type, 6 belonging to class III and 3 each to class IV, V and VI of the table. Also, it is easy to check that the structures of these 15 matrices correspond to either of the 15 matrices presented earlier, i.e., ${S_1}$ to $S_{15}$. Before proceeding further, it may be noted that the above mentioned patterns of texture 2 zero mass matrices have been deduced from 12 possible texture 3 zero patterns, the remaining 13th texture 3 zero pattern which is of the diagonal form, implying the three diagonal elements being non zero and all the off diagonal ones being zero, does not yield a texture 2 zero pattern.
 
%\begin{landscape}
\begin{sidewaystable}
\scriptsize

\begin{tabular}{|c|c|c|c|c|c|c|}
	\hline 
	&\text{Class I} &\text{Class II} & \text{Class III} & \text{Class IV} & \text{Class V} & \text{Class VI}  \\  
	\hline
a	& $\left(
\begin{array}{ccc}
	0 & A e^{\text{i$\alpha $}} & 0 \\
	A e^{-\text{i$\alpha $}} & 0 & B e^{\text{i$\beta $}} \\ 
	0 & B e^{-\text{i$\beta $}} & C
\end{array}
\right)$ &  $\left(
\begin{array}{ccc}
	0 & A e^{\text{i$\alpha $}} & 0 \\
	A e^{-\text{i$\alpha $}} & D & 0 \\
	0 & 0 & C
\end{array}
\right)$ & $\left(
\begin{array}{ccc}
	0 & A e^{\text{i$\alpha $}} & 0 \\
	A e^{-\text{i$\alpha $}} & D & B e^{\text{i$\beta $}} \\ 
	0 & B e^{-\text{i$\beta $}} & C
\end{array}
\right)$ & $\left(
\begin{array}{ccc}
	E & A e^{\text{i$\alpha $}} & 0 \\
	A e^{-\text{i$\alpha $}} & 0 & B e^{\text{i$\beta $}} \\
	0 & B e^{-\text{i$\beta $}} & C
\end{array}
\right)$  & $\left(
\begin{array}{ccc}
	0 & A e^{\text{i$\alpha $}} & F e^{\text{i$\gamma $}} \\
	A e^{-\text{i$\alpha $}} & 0 & B e^{\text{i$\beta $}} \\
	F e^{-\text{i$\gamma $}} & B e^{-\text{i$\beta $}} & C
\end{array}
\right)$  & $\left(
\begin{array}{ccc}
	E & A e^{\text{i$\alpha $}} & 0 \\
	A e^{-\text{i$\alpha $}} & D & 0 \\
	0 & 0 & C
\end{array}
\right)$ \\ 
\hline 
b	& $\left(
\begin{array}{ccc}
0 & 0 & A e^{\text{i$\alpha $}} \\
0 & C & B e^{-\text{i$\beta $}} \\
A e^{-\text{i$\alpha $}} & B e^{\text{i$\beta $}} & 0
\end{array}
\right)$ & $\left(
\begin{array}{ccc}
0 & 0 & A e^{\text{i$\alpha $}} \\
0 & C & 0 \\
A e^{-\text{i$\alpha $}} & 0 & D
\end{array}
\right)$& $\left(
\begin{array}{ccc}
0 & 0 & A e^{\text{i$\alpha $}} \\
0 & C & B e^{-\text{i$\beta $}} \\
A e^{-\text{i$\alpha $}} & B e^{\text{i$\beta $}} & D
\end{array}
\right)$ & $\left(
\begin{array}{ccc}
	E & 0 &  A e^{\text{i$\alpha $}} \\
	0 & C & B e^{-\text{i$\beta $}} \\
	 A e^{-\text{i$\alpha $}} & B e^{\text{i$\beta $}} & 0
\end{array}
\right)$  & $\left(
\begin{array}{ccc}
	0 & F e^{\text{i$\gamma $}} & A e^{\text{i$\alpha $}} \\
	F e^{-\text{i$\gamma $}} & C & B e^{-\text{i$\beta $}} \\
	A e^{-\text{i$\alpha $}} & B e^{\text{i$\beta $}} & 0
\end{array}
\right)$ & $\left(
\begin{array}{ccc}
E & 0 & A e^{\text{i$\alpha $}} \\
0 & C & 0 \\
A e^{-\text{i$\alpha $}} & 0 & D
\end{array}
\right)$ \\ 
	\hline 
c	& $\left(
\begin{array}{ccc}
0 & A e^{-\text{i$\alpha $}} & B e^{\text{i$\beta $}} \\
A e^{\text{i$\alpha $}} & 0 & 0 \\
B e^{-\text{i$\beta $}} & 0 & C
\end{array}
\right)$& $\left(
\begin{array}{ccc}
D & A e^{-\text{i$\alpha $}} & 0 \\
A e^{\text{i$\alpha $}} & 0 & 0 \\
0 & 0 & C
\end{array}
\right)$&$\left(
\begin{array}{ccc}
D & A e^{-\text{i$\alpha $}} & B e^{\text{i$\beta $}} \\
A e^{\text{i$\alpha $}} & 0 & 0 \\
B e^{-\text{i$\beta $}} & 0 & C
\end{array}
\right)$ & $\left(
\begin{array}{ccc}
	0 &  A e^{-\text{i$\alpha $}} & B e^{\text{i$\beta $}} \\
	 A e^{\text{i$\alpha $}} & E & 0 \\
	 B e^{-\text{i$\beta $}} & 0 & C
\end{array}
\right)$  & $\left(
\begin{array}{ccc}
	0 & A e^{-\text{i$\alpha $}} & B e^{\text{i$\beta $}} \\
	A e^{\text{i$\alpha $}} & 0 & F e^{\text{i$\gamma $}} \\
	B e^{-\text{i$\beta $}} & F e^{-\text{i$\gamma $}} & C
\end{array}
\right)$ & $\left(
\begin{array}{ccc}
D & A e^{-\text{i$\alpha $}} & 0 \\
A e^{\text{i$\alpha $}} & E & 0 \\
0 & 0 & C
\end{array}
\right)$ \\ 
	\hline 
d	& $\left(
\begin{array}{ccc}
C & B e^{-\text{i$\beta $}} & 0 \\
B e^{\text{i$\beta $}} & 0 & A e^{-\text{i$\alpha $}} \\
0 & A e^{\text{i$\alpha $}} & 0
\end{array}
\right)$ &   $\left(
\begin{array}{ccc}
C & 0 & 0 \\
0 & D & A e^{-\text{i$\alpha $}} \\
0 & A e^{\text{i$\alpha $}} & 0
\end{array}
\right)$& $\left(
\begin{array}{ccc}
C & B e^{-\text{i$\beta $}} & 0 \\
B e^{\text{i$\beta $}} & D & A e^{-\text{i$\alpha $}} \\
0 & A e^{\text{i$\alpha $}} & 0
\end{array}
\right)$  & $\left(
\begin{array}{ccc}
	C & B e^{-\text{i$\beta $}} & 0 \\
	B e^{\text{i$\beta $}} & 0 & A e^{-\text{i$\alpha $}}\\
	0 & A e^{\text{i$\alpha $}} & E
\end{array}
\right)$  & $\left(
\begin{array}{ccc}
	C & B e^{-\text{i$\beta $}} & F e^{-\text{i$\gamma $}} \\
	B e^{\text{i$\beta $}}& 0 & A e^{-\text{i$\alpha $}}  \\
	F e^{\text{i$\gamma $}} &  A e^{\text{i$\alpha $}} & 0
\end{array}
\right)$ &   $\left(
\begin{array}{ccc}
C & 0 & 0 \\
0 & D & A e^{-\text{i$\alpha $}} \\
0 & A e^{\text{i$\alpha $}} & E
\end{array}
\right)$  \\ 
	\hline 
e	& $\left(
\begin{array}{ccc}
0 & B e^{\text{i$\beta $}} & A e^{-\text{i$\alpha $}} \\
B e^{-\text{i$\beta $}} & C & 0 \\
A e^{\text{i$\alpha $}} & 0 & 0
\end{array}
\right)$& $\left(
\begin{array}{ccc}
D & 0 & A e^{-\text{i$\alpha $}} \\
0 & C & 0 \\
A e^{\text{i$\alpha $}} & 0 & 0
\end{array}
\right)$& $\left(
\begin{array}{ccc}
D & B e^{\text{i$\beta $}} & A e^{-\text{i$\alpha $}} \\
B e^{-\text{i$\beta $}} & C & 0 \\
A e^{\text{i$\alpha $}} & 0 & 0
\end{array}
\right)$ & $\left(
\begin{array}{ccc}
	0 & B e^{\text{i$\beta $}} & A e^{-\text{i$\alpha $}} \\
	B e^{-\text{i$\beta $}} & C & 0 \\
	A e^{\text{i$\alpha $}} & 0 & E
\end{array}
\right)$  & $\left(
\begin{array}{ccc}
	0 & B e^{\text{i$\beta $}} & A e^{-\text{i$\alpha $}} \\
	B e^{-\text{i$\beta $}} & C & F e^{-\text{i$\gamma $}} \\
	 A e^{\text{i$\alpha $}} & F e^{-\text{i$\gamma $}} & 0
\end{array}
\right)$ & $\left(
\begin{array}{ccc}
D & 0 & A e^{-\text{i$\alpha $}} \\
0 & C & 0 \\
A e^{\text{i$\alpha $}} & 0 & E
\end{array}
\right)$ \\ 
	\hline 
f	& $\left(
\begin{array}{ccc}
C & 0 & B e^{-\text{i$\beta $}} \\
0 & 0 & A e^{\text{i$\alpha $}} \\
B e^{\text{i$\beta $}} & A e^{-\text{i$\alpha $}} & 0
\end{array}
\right) $& $\left(
\begin{array}{ccc}
C & 0 & 0 \\
0 & 0 & A e^{\text{i$\alpha $}} \\
0 & A e^{-\text{i$\alpha $}} & D
\end{array}
\right)$&$\left(
\begin{array}{ccc}
C & 0 & B e^{-\text{i$\beta $}} \\
0 & 0 & A e^{\text{i$\alpha $}} \\
B e^{\text{i$\beta $}} & A e^{-\text{i$\alpha $}} & D
\end{array}
\right) $ & $\left(
\begin{array}{ccc}
	C & 0 & B e^{-\text{i$\beta $}} \\
	0 & E & A e^{\text{i$\alpha $}} \\
	B e^{\text{i$\beta $}} & A e^{-\text{i$\alpha $}} & 0
\end{array}
\right)$  & $\left(
\begin{array}{ccc}
	C & F e^{-\text{i$\gamma $}} & B e^{-\text{i$\beta $}} \\
	F e^{\text{i$\gamma $}} & 0 & A e^{\text{i$\alpha $}} \\
	B e^{\text{i$\beta $}} & A e^{-\text{i$\alpha $}} & 0
\end{array}
\right)$ & $\left(
\begin{array}{ccc}
C & 0 & 0 \\
0 & E & A e^{\text{i$\alpha $}} \\
0 & A e^{-\text{i$\alpha $}} & D
\end{array}
\right)$ \\ 
	\hline 
\end{tabular} 
\vspace{20pt}
\caption{Classes I and II: Possible texture 3 mass matrices; Classes III, IV ,V and VI: Possible texture 2 mass matrices} 
\label{table1}
\end{sidewaystable}
%\end{landscape}
  
Following the methodology presented in Ref. \cite{tex6zeroptep}, one may note that it essentially involves considering a possible texture 5 zero combination, i.e., $M_{U}$ being texture 3 zero type from either the 12 patterns listed in classes I and II of the table or a 13th pattern which is of the diagonal form, and  $M_{D}$ being texture 2 zero type , i.e., 15 independent patterns listed in classes III, IV, V and VI or vice versa. The viability of the considered combination is explored by examining the compatibility of the CKM matrix, constructed from a given combination of mass matrices, with the recent one given by Particle Data Group (PDG) \cite{pdg2018}. In particular, the CKM matrix can be obtained using the relation
\begin{equation}
V_{CKM}=O_{U}^{T}P_{U}P_{D}^{\dag}O_{D}= V_{U}^{\dag}V_{D},
\end{equation}
where the unitary matrices $V_{U}(= P_{U}^{\dag}O_{U})$ and
$V_{D}(=P_{D}^{\dag}O_{D})$ are the diagonalizing transformations for the
matrices $M_{U}$ and $M_{D}$ respectively. Here, $O_{U}$ and $O_{D}$ denote the orthogonal transformations for the real mass matrices $M_{i}^r$ ($i=U, D$) obtained after separating the phase matrices $P_{U}$ and $P_{D}$ from the respective mass matrices of the up and down sector. 
 In order to obtain $O_{U}$ and $O_{D}$ for the matrices belonging to class III, we consider the invariants trace $M_{i}^r$, trace $M_{i}^{r^{2}}$ and determinant $M_{i}^r$ to yield relations involving elements of the mass matrices. For all the six matrices belonging to class III of the table, using these invariants, we get
\begin{equation}
C_{i}=m_{1}-m_{2}+m_{3}-D_{i},~~~~|A_{i}|^2+|B_{i}|^2-C_{i}D_{i}=m_1m_2+m_2m_3-m_1m_3,~~~~|A_{i}|^2C_{i}=m_1m_2m_3,
\end{equation}
where the subscripts 1, 2 and 3 refer respectively to u, c
and t for the up sector and d, s and b for the down sector.
For matrix III$_{a}$, the corresponding  real matrix $ M_{i}^{r}$ can be expressed as 
\begin{equation}
M_{i}^{r}  =\begin{pmatrix}
  0 & |A_{i}| & 0 \\
  |A_{i}| & D_{i} & |B_{i}| \\
  0 & |B_{i}| & C_{i} \\
\end{pmatrix}
\end{equation}
and $P_{i}$, the phase matrix, is given by
\begin{equation}
    P_{i}=\begin{pmatrix}
      e^{-\text{i$\alpha $}_i} & 0 & 0 \\
      0 & 1 & 0 \\
      0 & 0 & e^{\text{i$\beta $}_i} \\
    \end{pmatrix}.
\end{equation}
Corresponding to the matrix III$_{a}$, the diagonalizing transformation $O_{i}$ is given as 
\begin{equation}
O_i\text{  }=\text{     }\left(
\begin{array}{ccc}
	\sqrt{\frac{m_2m_3\left(m_3-m_2\right)}{C_i\left(m_1+m_2\right)\left(m_3-m_1\right)}} & \sqrt{\frac{m_1m_3\left(m_1+m_3\right)}{C_i\left(m_1+m_2\right)\left(m_3+m_2\right)}} & \sqrt{\frac{m_1m_2\left(m_2-m_1\right)}{C_i\left(m_3+m_2\right)\left(m_3-m_1\right)}} \\
	\sqrt{\frac{m_1\left(m_3-m_2\right)}{\left(m_1+m_2\right)\left(m_3-m_1\right)}} & -\sqrt{\frac{m_2\left(m_1+m_3\right)}{\left(m_1+m_2\right)\left(m_3+m_2\right)}} & \sqrt{\frac{m_3\left(m_2-m_1\right)}{\left(m_3-m_1\right)\left(m_2+m_3\right)}} \\
	-\sqrt{\frac{m_1\left(m_1+m_3\right)\left(m_2-m_1\right)}{C_i\left(m_1+m_2\right)\left(m_3-m_1\right)}} & \sqrt{\frac{m_2\left(m_3-m_2\right)\left(m_2-m_1\right)}{C_i\left(m_1+m_2\right)\left(m_3+m_2\right)}} & \sqrt{\frac{m_3\left(m_3-m_2\right)\left(m_3+m_1\right)}{C_i\left(m_3+m_2\right)\left(m_3-m_1\right)}}
\end{array}
\right).	
\end{equation}	
It may be mentioned that the rest of the 5 matrices belonging to class III of the table can be similarly expressed in terms of a real matrix $ M_{i}^{r}$ and the corresponding phase matrix $P_{i}$, yeilding the corresponding $O_{i}$ in a similar manner. Corresponding details regarding the matrices belonging to classes I and II have been presented in Ref \cite{tex6zeroptep}.

To arrive at texture 5 zero combinations, as a first step, let us consider the matrices belonging to class III of the table to be either $M_{U}$ or $M_{D}$. Along with this, we can consider the matrices for the other sector to be belonging to either class I or class II. As a result, one arrives at the following possible texture 5 zero combinations:
\\
\\
Category 1: $M_{U}$ from class III and $M_{D}$ from class I.\\
Category 2: $M_{U}$ from class III and $M_{D}$ from class II.\\
Category 3: $M_{U}$ from class I and $M_{D}$ from class III.\\
Category 4: $M_{U}$ from class II and $M_{D}$ from class III.\\
 
To begin with , we consider Category 1, wherein the matrices $M_{U}$ and $M_{D}$ can be either of the 6 matrices III$_{a-f}$ and I$_{a-f}$ respectively, therefore, yielding a total of 36 combinations. We first consider the combinations III$_a$I$_a$, III$_b$I$_b$, etc., wherein, the texture 2 zero matrices of class III are very similar to the corresponding matrices of class I with one of their zeros being replaced by the non zero element D. For the purpose of numerical analysis, we consider the ``current'' quark masses at $M_{Z}$ energy scale \cite{masses}  given by
\begin{eqnarray}
m_{u}=1.45_{-0.45}^{+0.56}~{\rm MeV},~~~m_{d}=2.9_{-0.4}^{+0.5}~{\rm MeV}, 
~~~m_{s}=57.7_{-15.7}^{+16.8}~{\rm MeV},~~~~~~~~~~~  \nonumber 
\\
m_{c}= 0.635 \pm 0.086 ~{\rm GeV},~~~ m_{b}=2.82_{-0.04}^{+0.09}~ {\rm GeV},~~~ m_{t}=172.1\pm 0.6\pm 0.9 ~{\rm GeV}. 
\label{inputs}
\end{eqnarray} 
The quark mass ratios $\frac{m_u}{m_d}$ and $\frac{m_s}{m_{ud}}$, wherein ${m_{ud}}$ is defined as $\frac{1}{2(m_u + m_d)}$ are given by \cite{flag}
\begin{equation}
\frac{m_u}{m_d}= 0.45~(3)~~~~~{\rm and}~~~~~\frac{m_s}{m_{ud}}= 27.30~(34).
\label{massrat}
\end{equation}
In the absence of any information related to the phases
associated with the elements of the mass matrices $\phi_{1}=\alpha_{U}-\alpha_{D}$ and
$\phi_{2}=\beta_{U}-\beta_{D}$, these have been given full variation from $0^o$ to $360^o$. Further, for the mass matrices to remain `natural' \cite{ourrevart2}, the parameter $D_U$ of the mass matrix $M_U$ has been restricted within the range $0 - m_t$ or $0-172$ GeV. Along with these inputs, to facilitate discussion, we have imposed the recent values of well known CKM matrix elements $V_{us}$ and $V_{ub}$ as per PDG 2018 \cite{pdg2018}  as a constraint
\begin{equation}
V_{us}= 0.2243 \pm 0.0005,~~~~~~~~~~~V_{ub}= (3.94 \pm 0.36) \times 10^{-3}.
\label{cons}
\end{equation}

Before proceeding further, it is perhaps desirable to mention that the complex unitary CKM matrix can be characterized by three real mixing angles and one CP violating phase, equivalently, it can also be characterized by magnitudes of three CKM matrix elements and a CP violating parameter or by magnitudes of the elements of the CKM matrix \cite{fxrevart, ourrevart2}. In the present case, we compare our predictions with the magnitudes of the CKM matrix elements and CP violating phase represented by Jarlskog’s rephasing invariant parameter $J$.

Using the above mentioned inputs and constraints, the corresponding CKM matrices for all the 6 combinations III$_a$I$_a$, III$_b$I$_b$, etc., come out to be the same, i.e.,
\begin{equation}
V_{CKM}=\begin{pmatrix}
  0.9743-0.9746 & 0.2238-0.2247 & 0.0036-0.0042 \\
  0.2227-0.2242 & 0.9685-0.9727 & 0.0608-0.1099 \\
  0.0131-0.0241 & 0.0595-0.1074 & 0.9939-0.9981 \\
\end{pmatrix}.
\end{equation}
As mentioned earlier, the above matrix presents the numerical ranges of the magnitudes of the CKM matrix elements. A look at this matrix immediately reveals that the ranges of some of the CKM matrix elements show no overlap with those obtained by recent global analysis  \cite{pdg2018}, 
\begin{equation}
    V_{CKM}=\begin{pmatrix}
      0.9744-0.9746 & 0.2241-0.2250 & 0.0035-0.0038 \\
      0.2239-0.2248 & 0.9735-0.9737 & 0.0414-0.0429 \\
      0.0087-0.0092 & 0.0406-0.0421 & 0.9990-0.9991 \\
    \end{pmatrix}.
    \label{pdgmatrix}
\end{equation}
Further, we have also evaluated the CP asymmetry parameter $Sin2\beta$ and the Jarlskog's rephasing invariant parameter $J$, which come out to be 
\begin{equation}
Sin2\beta= 0.311-0.511,~~~J =(4.73-9.92)\times 10^{-5},
 \end{equation}
again showing no overlap with their experimentally determined ranges \cite{pdg2018} 
\begin{equation}
Sin2\beta= 0.674-0.708,~~~J =(3.03-3.33)\times 10^{-5}.  
\end{equation}

From the above discussion, one can conclude that the texture combination III$_a$I$_a$ as well as the other similar combinations III$_b$I$_b$, etc., are essentially ruled out. Further, it is interesting to examine to what extent these combinations remain ruled out. To this end, in Figure 1, we have presented the dependence of CKM matrix element $V_{cb}$, CP asymmetry parameter $Sin2\beta$ and Jarlskog's rephasing invariant parameter $J$ w.r.t. the light quark mass $m_u$. While plotting these graphs, we have considered a relatively wider range of mass $m_u$, i.e., from $0-3.0$ ${\rm MeV}$. The vertical lines in these plots depict the range of $m_u$ given in equation (\ref{inputs}), whereas the horizontal lines show the experimental ranges \cite{pdg2018} of the matrix elements $V_{cb}$ and the parameters $Sin2\beta$ and $J$. These plots reveal that corresponding to the experimental range of $m_u$, the allowed ranges of $V_{cb}$, $Sin2\beta$ and $J$ obtained here show essentially no overlap with their experimentally determined ranges, however, with the possibility of a slight overlap in case the upper limit of $m_u$ gets pushed slightly higher.
\begin{figure}
\begin{multicols}{3}
{\hspace*{-45pt}\includegraphics[width=8cm,height=6cm]{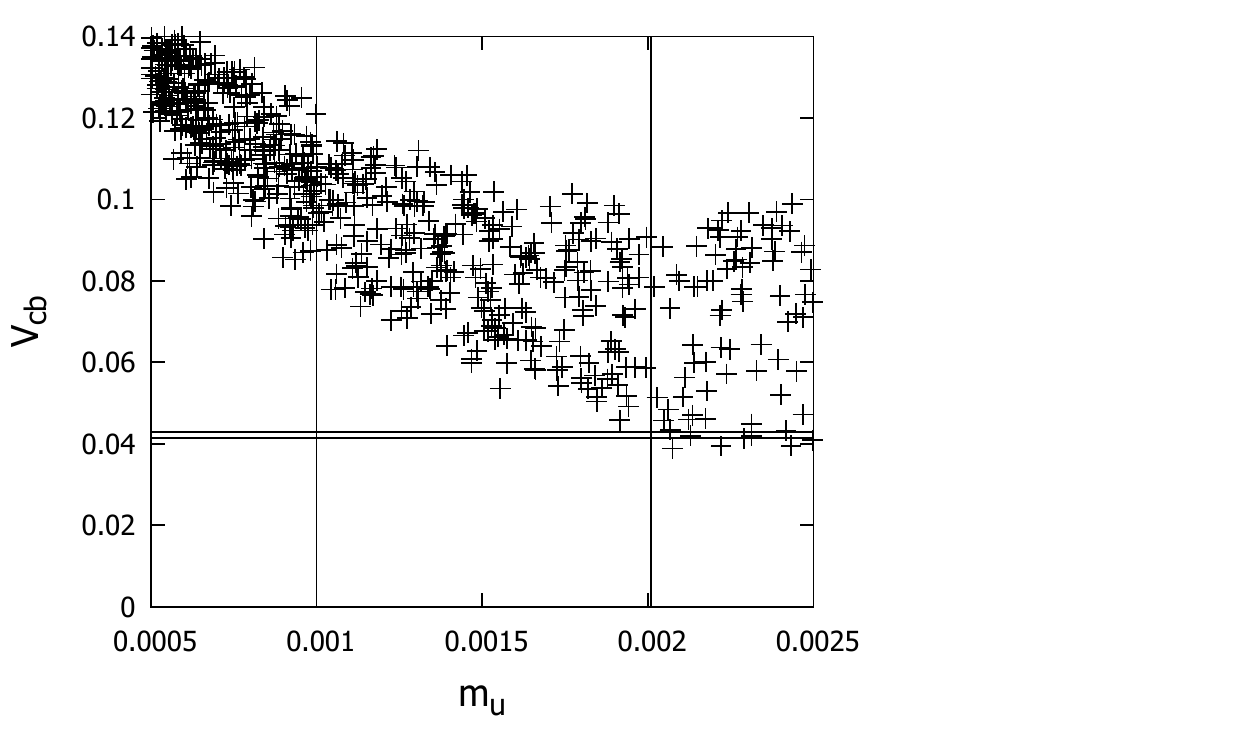}} 
{\hspace*{-8pt}\includegraphics[width=8cm,height=6cm]{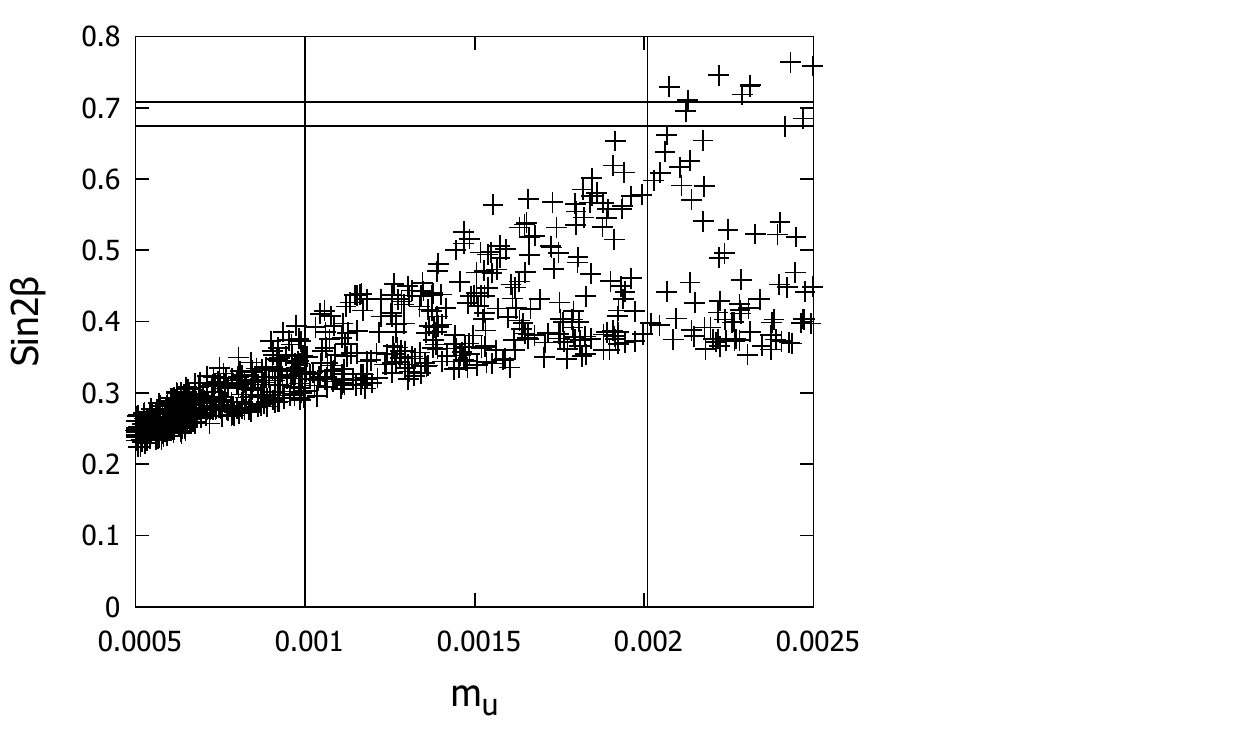}} 
{ \includegraphics[width=8cm,height=6cm]{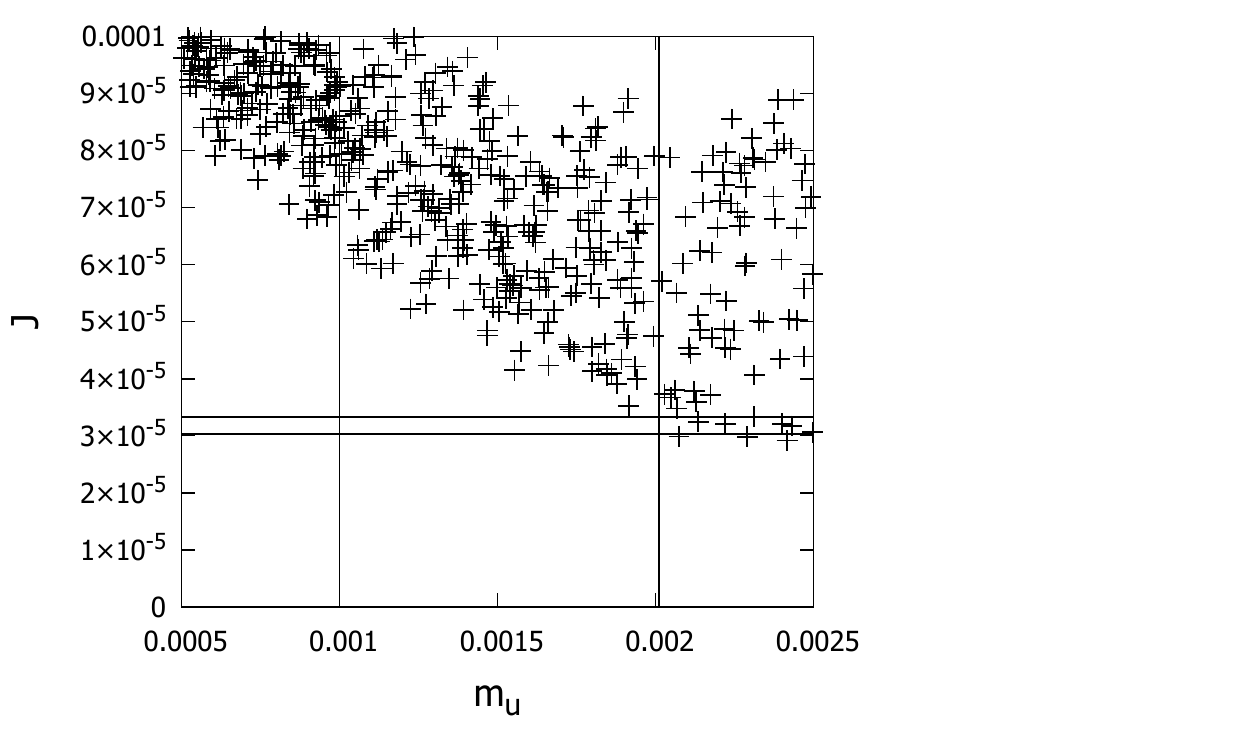}}
\end{multicols} \vspace*{-15pt}
Figure 1:  Allowed ranges of $V_{cb}$, $Sin2\beta$ and $J$ w.r.t the light quark mass $m_u$ (\rm {in GeV})
\label{figr1}
\end{figure}

In order to investigate this further, in Figures 2 and 3, we have again examined the variation of these 3 parameters w.r.t the other light quark masses $m_d$ and $m_s$ as well. Again, considering mass $m_u$ from $0-3.0$ ${\rm MeV}$, we have obtained wider ranges of masses $m_d$ and $m_s$ using the mass ratios mentioned in equation (\ref{massrat}). A look at these plots reveal that the allowed ranges of these parameters clearly do not show any overlap with the experimental results even if, in future, the ranges of the quark masses $m_d$ and $m_s$ become considerably wider, re-emphasising the earlier conclusion regarding the ruling out of these combinations of mass matrices. 

\begin{figure}
\begin{multicols}{3}
{\hspace*{-45pt}\includegraphics[width=8cm,height=6cm]{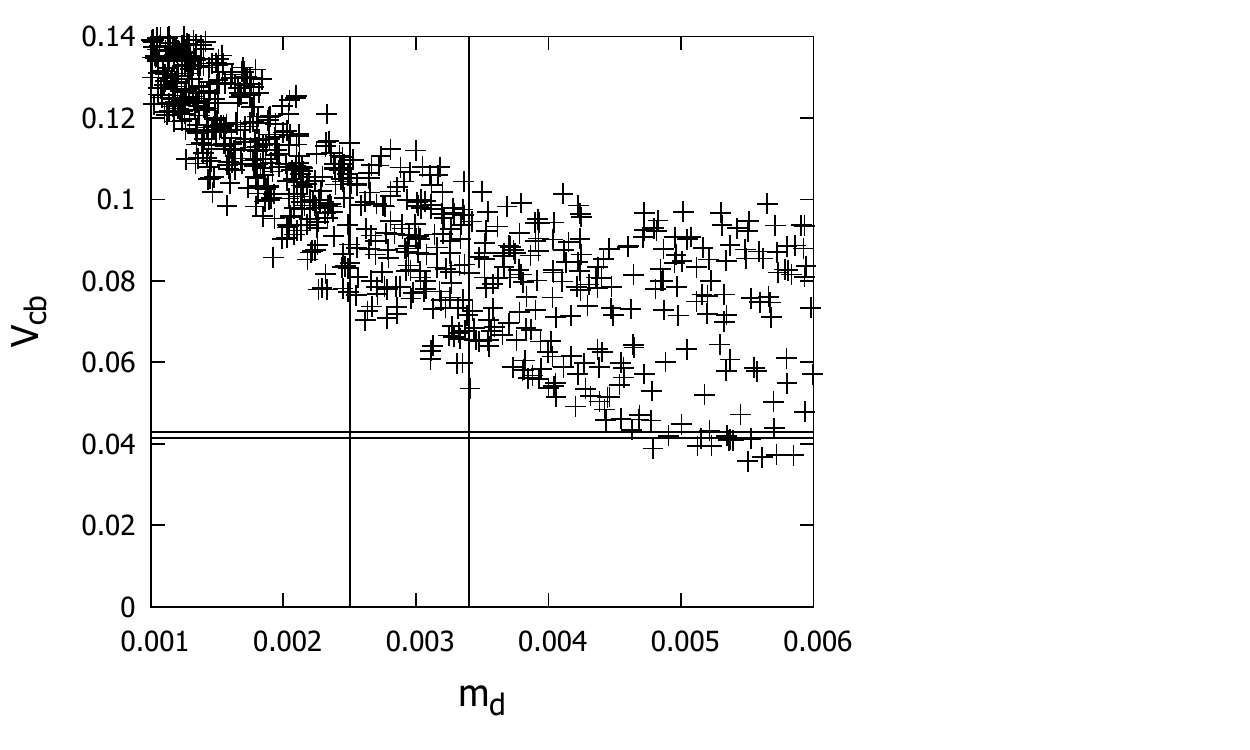}} 
{\hspace*{-8pt}\includegraphics[width=8cm,height=6cm]{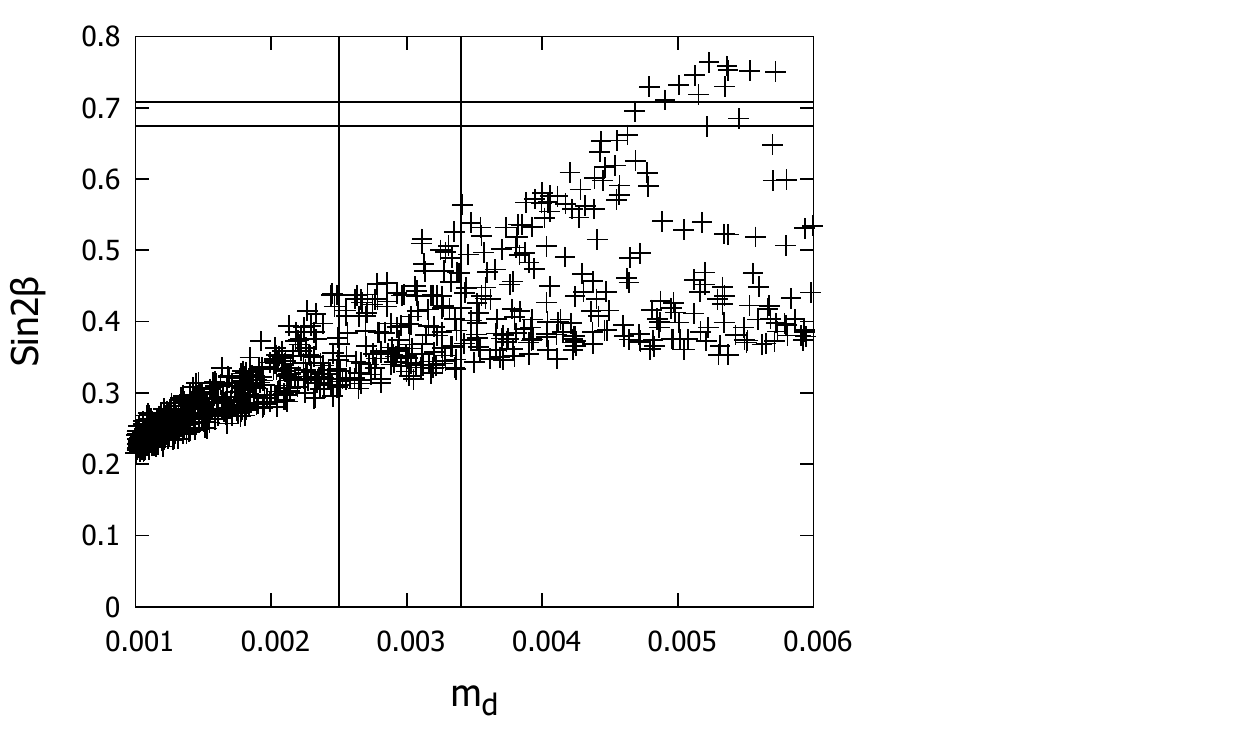}} 
{ \includegraphics[width=8cm,height=6cm]{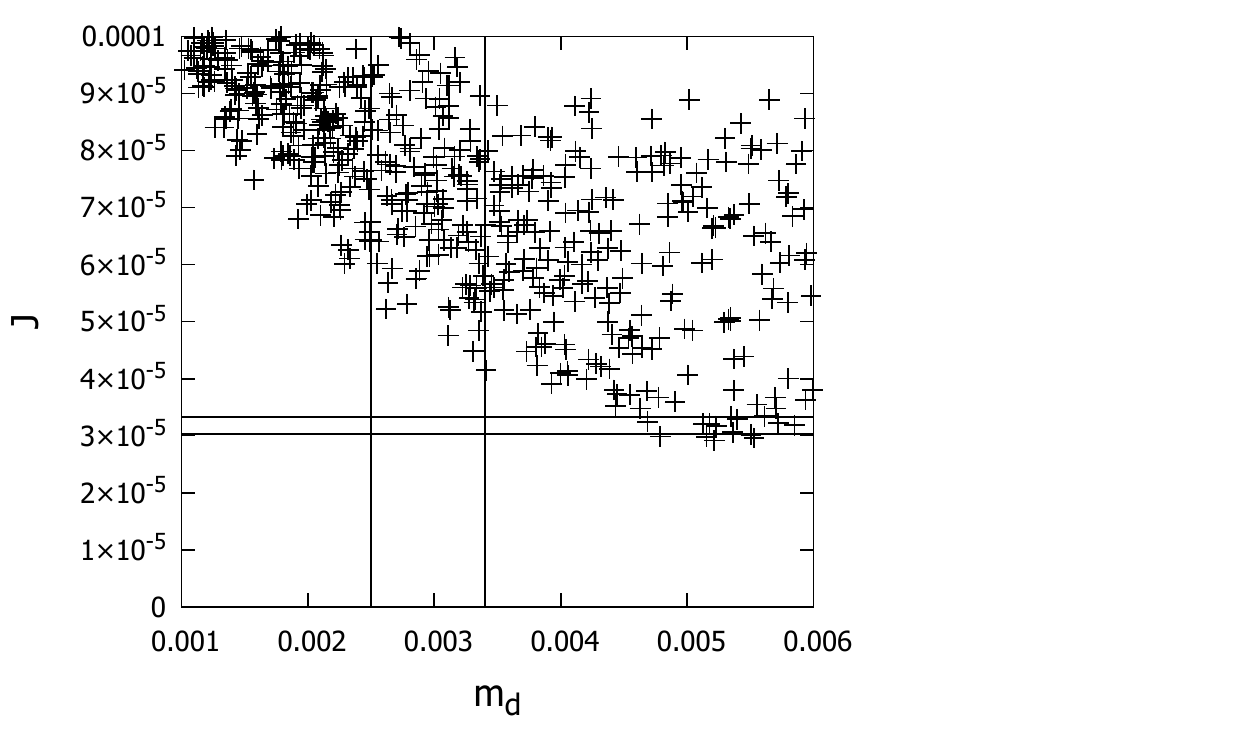}}
\end{multicols} \vspace*{-15pt}
Figure 2: Allowed ranges of $V_{cb}$, $Sin2\beta$ and $J$ w.r.t the light quark mass $m_d$ (\rm {in GeV})
\label{figr2}
\end{figure}

\begin{figure}
\begin{multicols}{3}
{\hspace*{-45pt}\includegraphics[width=8cm,height=6cm]{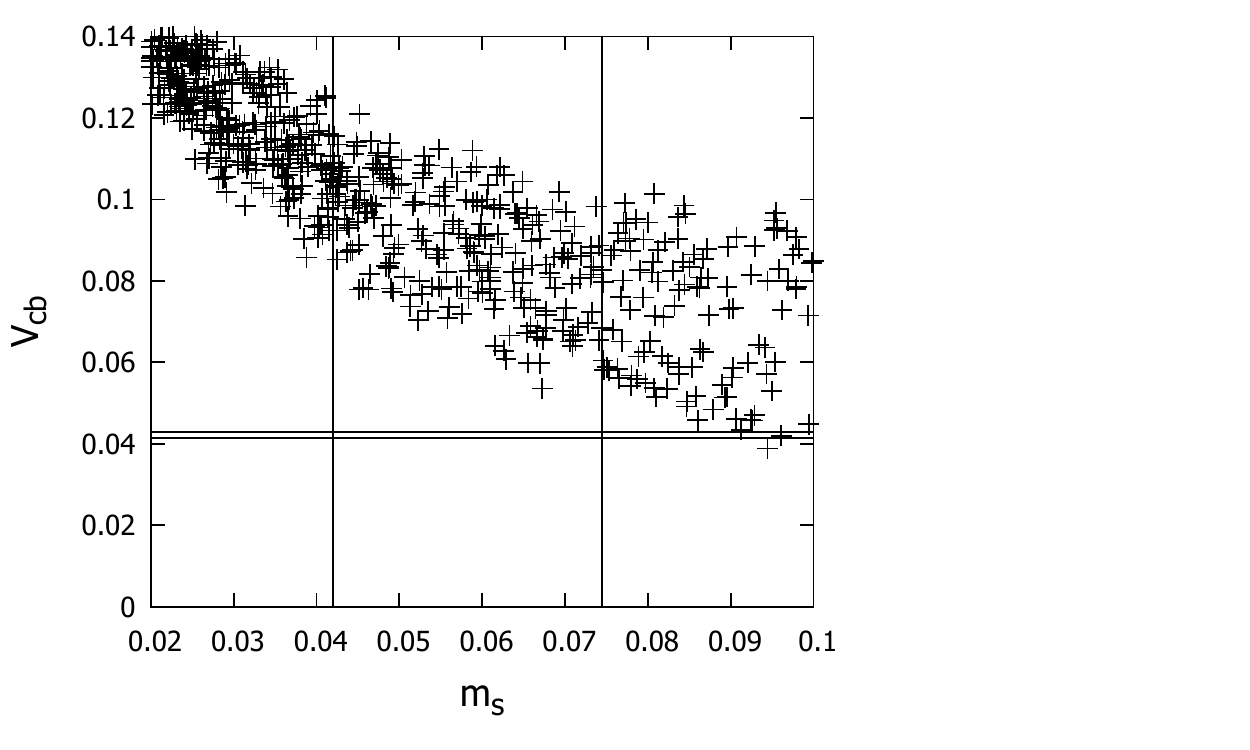}} 
{\hspace*{-8pt}\includegraphics[width=8cm,height=6cm]{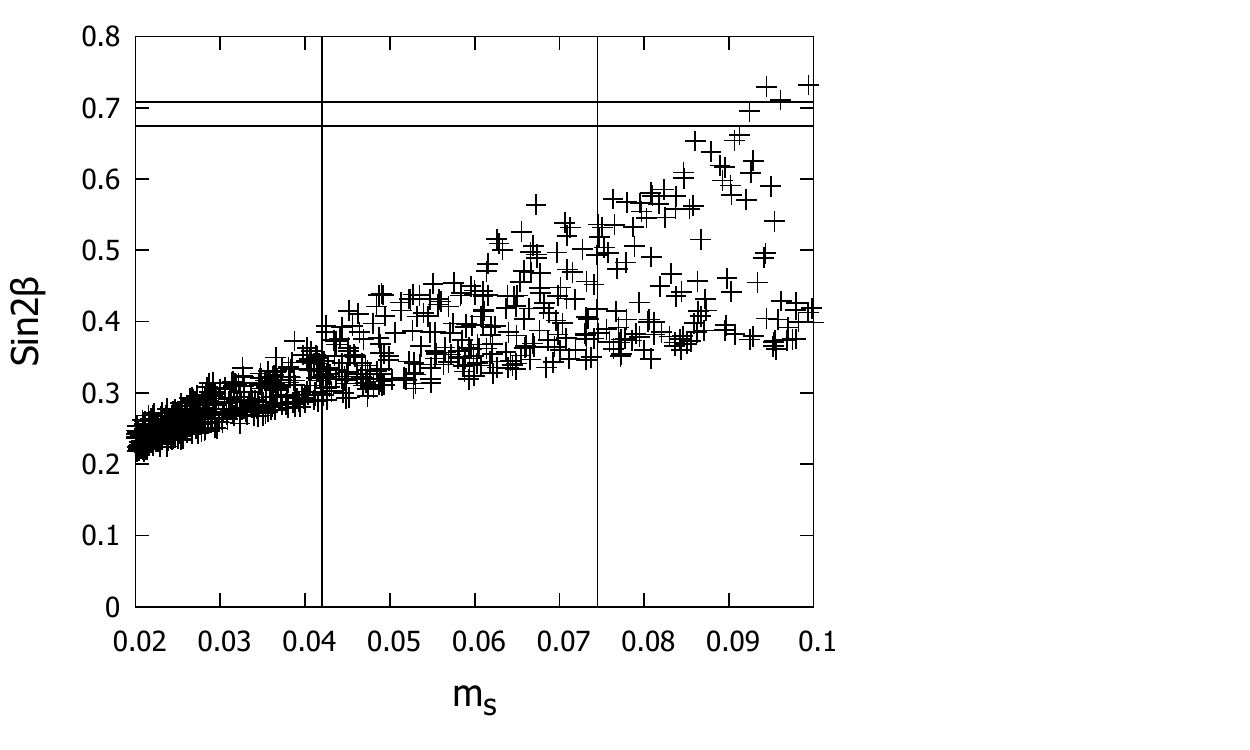}} 
{ \includegraphics[width=8cm,height=6cm]{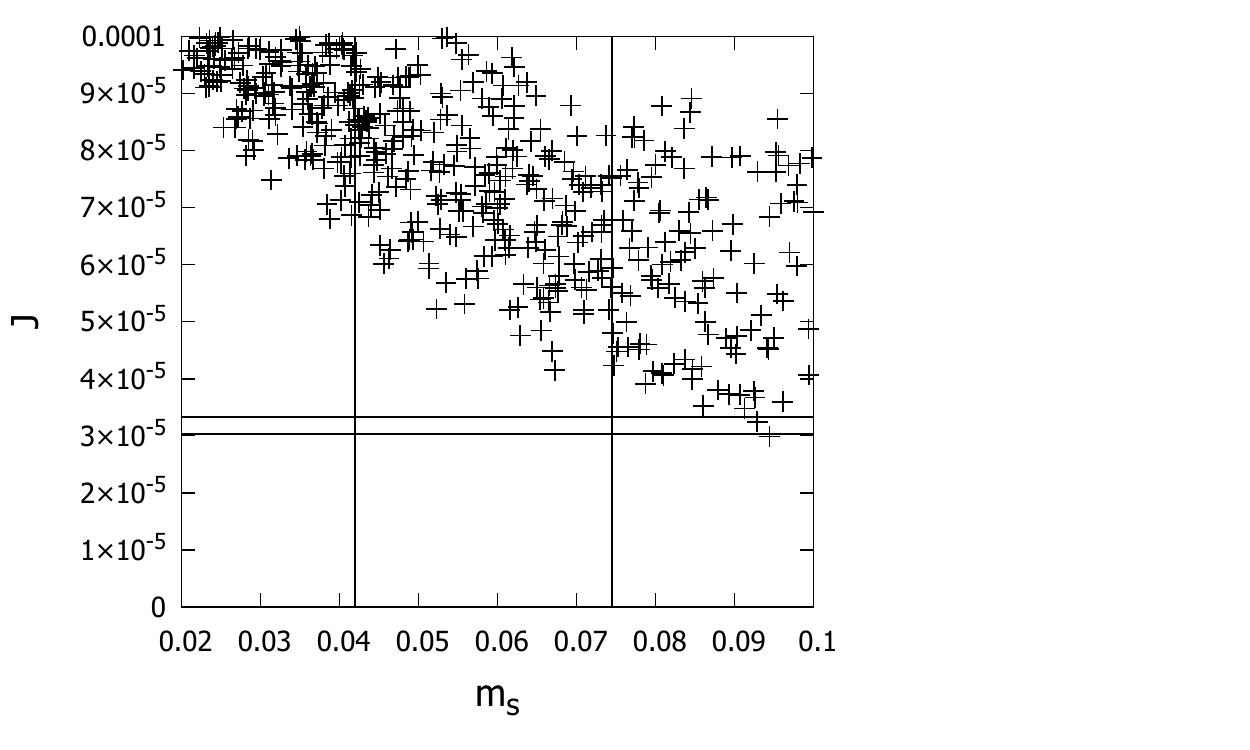}}
\end{multicols} \vspace*{-15pt}
Figure 3: Allowed ranges of $V_{cb}$, $Sin2\beta$ and $J$ w.r.t the light quark mass $m_s$ (\rm {in GeV})
\label{figr3}
\end{figure}

Further, for these texture combinations III$_a$I$_a$, III$_b$I$_b$, etc., wherein $M_U$ is considered to be a texture 2 zero mass matrix, it may be of interest to discuss the role of the parameter $D_U$ of the mass matrix $M_U$, which is absent in case $M_U$ is considered to be a texture 3 zero mass matrix. In this context, in Figure 4, we have presented the plot of the CKM matrix element $V_{cb}$ w.r.t. the parameter $D_U$. It may be noted that while carrying out the analysis, $D_U$ has been provided the earlier mentioned full variation from $0-m_t$ or $0-172$ GeV, however, the plot has been presented for $D_U$ upto 10 GeV only, the conclusions, however, remain the same for the entire range of $D_U$. A look at the graph immediately reveals that the allowed range of the element $V_{cb}$ shows no overlap with its experimentally determined range, shown as horizontal lines, therefore, leading to the conclusion that despite giving full variation to $D_U$, these combinations of the mass matrices remain ruled out.

\begin{figure}
\begin{center}
{\hspace*{+50pt}\includegraphics[width=8cm,height=6cm]{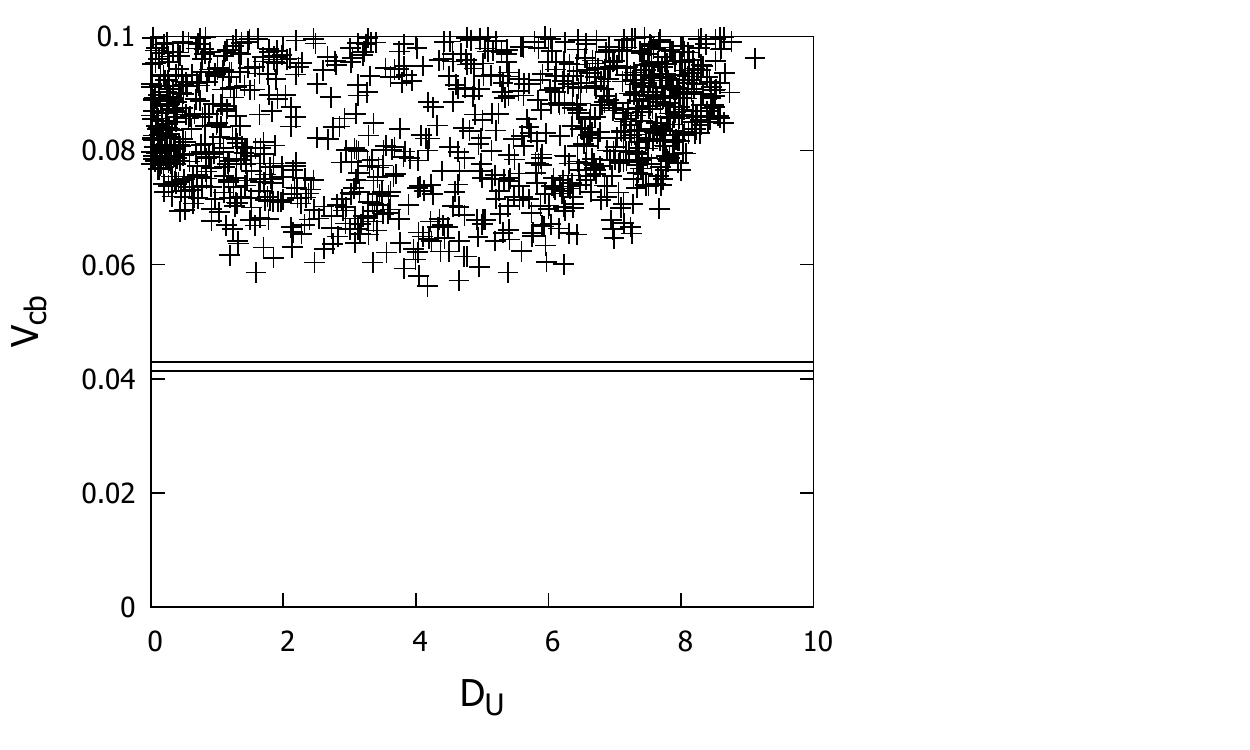}} 
\vspace*{-15pt}
\end{center}
Figure 4: Allowed range of $V_{cb}$ w.r.t. the mass matrix parameter $D_U$ (\rm {in GeV})
\label{figr4}
\end{figure}

For the remaining 30 combinations of Category 1, wherein $M_{U}$ and $M_{D}$ can be of the type III$_a$I$_b$, III$_c$I$_d$, etc., one can rule these out from the analytical expressions of the CKM matrix elements  alone. In particular, one finds that the CKM matrix does not have the usual structure wherein the diagonal elements are almost unity whereas the off diagonal elements are much smaller than these, thus, ruling out all these combinations. For the combination  III$_a$I$_b$, the leading order CKM matrix obtained is
\begin{equation}
V_{CKM}=\left(
\begin{array}{ccc}
	e^{\text{-i($\alpha $}_D+\text{$\alpha $}_U)} & \sqrt{\frac{m_d}{m_s}}  & \sqrt{\frac{m_u}{m_t}}e^{-\text{i$\beta $}_D} \\
	e^{\text{-i($\alpha $}_D+\text{$\alpha $}_U)}\sqrt{\frac{m_u}{m_c}} & -e^{-\text{i$\beta $}_U} \sqrt{\frac{m_c}{m_t}}-e^{\text{i$\beta $}_D}\sqrt{\frac{m_s}{m_b}} & -e^{-\text{i$\beta $}_D} \\
	e^{-\text{i$\beta $}_U}-^{\text{i$\beta $}_D} \sqrt{\frac{m_s}{m_b}}\sqrt{\frac{m_c}{m_t}} &-e^{-\text{i$\beta $}_U}+e^{\text{i$\beta $}_D} \sqrt{\frac{m_s}{m_b}}\sqrt{\frac{m_c}{m_t}} & e^{-\text{i$\beta $}_U} \sqrt{\frac{m_s}{m_b}}+e^{\text{i$\beta $}_D}\sqrt{\frac{m_c}{m_t}}
\end{array}
\right),
\end{equation}
its numerical analysis yielding
\begin{equation}
    V_{CKM}=\begin{pmatrix}
      0.1667-0.2664 & 0.9505-0.9779 & 0.0035-0.0042 \\
      0.0858-0.1989 & 0.0051-0.3092 & 0.0540-0.9766 \\
      0.0292-0.9609 & 0.0047-0.2631 & 0.1841-0.9881 \\
    \end{pmatrix},
\end{equation}
which, clearly, does not have the usual structure.   

As a next step, we proceed towards the discussion of Category 2, wherein $M_{U}$ is a matrix from class III of the table and $M_{D}$ is a matrix from class II. Interestingly, one finds results similar to those for combinations of Category 1, in particular, six cases III$_a$II$_a$, III$_b$II$_b$, etc., yield the following CKM matrix
\begin{equation}
    V_{CKM}=\begin{pmatrix}
      0.9743-0.9746 & 0.2238-0.2247 & 0.0035-0.0042 \\
      0.2228-0.2242 & 0.9690-0.9720 & 0.0715-0.1043 \\
      0.0157-0.0231 & 0.0669-0.1018 & 0.9945-0.9974 \\
    \end{pmatrix}.
\end{equation}
This matrix, showing no overlap of the ranges of some of the CKM matrix elements with those obtained through global analysis, mentioned in equation (\ref{pdgmatrix}), clearly indicates the ruling out of these combinations. The same can be checked by plots showing dependence of CKM matrix elements $V_{cs}$, $V_{cb}$, $V_{td}$, $V_{ts}$, $V_{tb}$, the Jarlskog's rephasing invariant parameter $J$ and CP asymmetry parameter $Sin2\beta$. These plots, not shown here, are quite similar to the corresponding ones given for Category 1. For the remaining 30 combinations of Category 2, similar to the results for Category 1, one finds these combinations being ruled out analytically alone. 

As a next step, we consider the combinations pertaining to Categories 3 and 4, wherein $M_{U}$ is a matrix from class I and II of the table respectively and $M_{D}$ is a matrix from class III. In this regard, the analysis has been carried out using the same inputs as for the above mentioned analyses of the combinations of the mass matrices pertaining to Categories 1 and 2, however, the parameter $D_D$ of the mass matrix $M_D$ has been restricted within the range  $0 - m_b$ or $0-3$ GeV. Interestingly, one finds that a similar analyses of these categories indicates the 36 combinations of each of these again being ruled out. This, therefore, leads to ruling out of all the combinations involving matrices of class III in all the 4 categories.

Before presenting the analyses pertaining to the matrices belonging to classes IV, V and VI, for the sake of completion, we discuss the case of texture 5 zero combinations obtained by considering texture 3 zero mass matrix to be of the diagonal form, i.e., having three non vanishing diagonal elements with the non diagonal ones being zero, and texture 2 zero mass matrix belonging to class III of the table. This leads to a total of 12 combinations, either $M_{U}$ being texture 3 zero diagonal matrix and $M_{D}$ being any one of the 6 matrices listed in class III or vice versa. Interestingly, one finds that all of these 12 combinations get ruled out by the existing data. For example, considering $M_{U}$ being texture 3 zero matrix having the diagonal form and $M_{D}$ being III$_a$, using the inputs and constraints mentioned in equation (\ref{inputs}), (\ref{massrat}) and (\ref{cons}) respectively, we obtain the following CKM matrix
\begin{equation}
    V_{CKM}=\begin{pmatrix}
      0.9743-0.9748 & 0.2238-0.2247 & 0.0035-0.0042 \\
      0.1462-0.1830 & 0.6504-0.8077 & 0.5601-0.7453 \\
      0.1283-0.1696 & 0.5452-0.7264 & 0.6650-0.8283 \\
    \end{pmatrix}.
\end{equation}
A look at the above matrix clearly reveals that the ranges of some of the CKM matrix elements show no overlap with those obtained through global analysis, given in equation (\ref{pdgmatrix}), hence, ruling out this combination of mass matrices.

Similarly, if we consider the other combination wherein $M_{U}$ is considered to be III$_a$ and $M_{D}$ to be texture 3 zero matrix having the diagonal form, one gets
\begin{equation}
    V_{CKM}=\begin{pmatrix}
      0.9985-0.9992 & 0.0373-0.0536 & 0.0025-0.0132 \\
      0.0384-0.0542 & 0.9672-0.9973 & 0.0564-0.2496 \\
      0.8\times{10^{-5}}-0.5\times{10^{-4}} & 0.0565-0.2498 & 0.9682-0.9983 \\
    \end{pmatrix}.
\end{equation}
Again, it can be seen that the above CKM matrix is not compatible with the CKM matrix obtained through global analysis, thereby, ruling out this combination of mass matrices as well. For the remaining 10 combinations, a similar analysis yields CKM matrices not having the usual structure, i.e., the diagonal elements being almost unity and the off diagonal ones being much smaller, therefore, ruling out all these 10 combinations.

After ruling out all the combinations involving matrices of class III, we now present the analysis pertaining to the matrices belonging to class IV. For all the matrices belonging to this class of the table, the relations of the mass matrix elements in terms of the quark masses are given by
\begin{equation}
C_{i}=m_{1}-m_{2}+m_{3}-E_{i},~~~~|A_{i}|^2+|B_{i}|^2-C_{i}E_{i}=m_1m_2+m_2m_3-m_1m_3,~~~~|A_{i}|^2C_{i}+|B_{i}|^2E_{i}=m_1m_2m_3.
\end{equation} 
To begin with, we consider the combination IV$_{a}$I$_{a}$, wherein $M_{U}$ is the first matrix from class IV of the table, corresponding to texture 2 zero mass matrix and $M_{D}$ is the first matrix from class I, corresponding to the texture 3 zero mass matrix . For the matrix IV$_{a}$, the corresponding  real matrix $ M_{i}^{r}$ can be expressed as
\begin{equation}
M_{i}^{r}=\begin{pmatrix}
  E_{i} & |A_{i}| & 0 \\
  |A_{i}| & 0 & |B_{i}| \\
  0 & |B_{i}|  & C_{i} \\
\end{pmatrix},
\end{equation}
whereas the phase matrix $P_i$ is given by
\begin{equation}
    P_{i}=\begin{pmatrix}
      e^{-\text{i$\alpha $}_i} & 0 & 0 \\
      0 & 1 & 0 \\
      0 & 0 & e^{\text{i$\beta $}_i} \\
    \end{pmatrix}.
\end{equation}
Considering the inputs and constraints  mentioned in equations (\ref{inputs}), (\ref{massrat}) and (\ref{cons}) respectively, an analysis similar to the one presented above for the matrices belonging to class III of the table yields the following CKM matrix
\begin{equation}
    V_{CKM}=\begin{pmatrix}
      0.9744-0.9746 & 0.2238-0.2247 & 0.0035-0.0042 \\
      0.3997-0.5167 & 0.9435-0.9999 & 0.9400-0.9998 \\
      0.0255-0.0355 & 0.1127-0.1543 & 0.9873-0.9932 \\
    \end{pmatrix}.
 \end{equation}
Interestingly, one finds that again the above matrix is not compatible with the one obtained through global analysis, mentioned in equation (\ref{pdgmatrix}), thereby, ruling out this combination of mass matrices. Before coming to the discussion of the other combinations belonging to this class, in order to have a better understanding of the structures of the mass matrices $M_{U}$ and $M_{D}$, it is perhaps desirable to examine the numerical values corresponding to these, given by
\begin{equation}
    M_U=\begin{pmatrix}
      0.0005-0.0009 & 0.0134-0.0213 & 0 \\
       0.0134-0.0213& 0 & 9.7093-11.1704 \\
       0 & 9.7093-11.1704 & 169.909-173.001 \\
    \end{pmatrix}
\end{equation}
and
\begin{equation}
    M_D=\begin{pmatrix}
      0 & 0.0114-0.0152 & 0 \\
      0.0114-0.0152& 0 & 0.3607-0.4273 \\
       0 & 0.3607-0.4273 & 2.7268-2.8597 \\
    \end{pmatrix}.
\end{equation}
Clearly, a look at the above matrix $M_{U}$ reveals that its 1,1 element is quite small in comparison with the other non zero elements, thereby, leading to the conclusion that it may essentially be considered to be effectively reducing to a texture 3 zero matrix only, analysis of which has already been carried out in detail \cite{tex6zeroptep}. Examining the structure of the mass matrices $M_{U}$ and $M_{D}$ corresponding to the other combinations of not only this class, but also for all the matrices of classes V and VI yield similar results, therefore, indicating that a case by case analysis for the various combinations of the matrices of these classes is not required. Also, pertaining to the matrices of these classes, the texture 5 zero combinations obtained by considering texture 3 zero mass matrix to be of the diagonal form and texture 2 zero mass matrix belonging to classes IV, V and VI of the table or vice versa, one again finds all these combinations being ruled out on lines similar to the earlier cases corresponding to matrices belonging to class III.

Before we conclude, it is pertinent to mention that mass matrices are formulated at GUT scale whereas their analysis is carried out at weak scale. One may wonder whether the texture structure of mass matrices may undergo change as one uses Renormalization Group (RG) equations to scale down from GUT scale to weak scale, essentially requiring change in the analysis carried out. However, it may be mentioned that it has been shown \cite{fxrevart,tsmmrrr} that texture structure is invariant with respect to the  RG evolution between these scales. In any case, we would like to emphasize that the texture zeros are not exact zeros, these are phenomenological zeros, therefore, even if there are changes imposed by RG equations these would be of minor nature without changing our analysis and conclusions. 

To summarize, keeping in mind, refinements in the measurements of light quark masses $m_u$, $m_d$ and $m_{s}$ as well as in the CKM matrix elements, we have carried out a detailed analysis of quark mass matrices having structure beyond the minimal texture, implying texture 6 zero quark mass matrices. In particular, we have examined large number of combinations of texture 5 zero quark mass matrices to check the compatibility of the corresponding CKM matrix, arrived through these, with the latest one given by PDG. Interestingly, we find that all these possibilities are excluded by the present quark mixing data and the improved limits of the light quark masses. Further, these conclusions would remain largely valid even if, in future, there are changes in the ranges of the light quark masses, having important implications for model building.
\\
\\
{\bf Acknowledgements} \\ The authors would like to thank Chairman, Department of Physics, P.U., for providing facilities to work.

\end{document}